\documentclass[11pt,twoside]{article}
\usepackage{asp2010}

\resetcounters

\begin{document}

\title{The DRAO Synthesis Telescope in the post-CGPS Era}
\author{R. Kothes, T. L. Landecker, and A. D. Gray
\affil{National Research Council Canada, Herzberg Institute of Astrophysics,
Dominion Radio Astrophysical Observatory, P.O. Box 248, Penticton, BC V2A~6J9, Canada}}

\begin{abstract}
The DRAO ST was used for the past 15 years as the primary instrument for the Canadian 
Galactic Plane Survey \citep[CGPS,][]{tayl03}. This has been a spectacularly 
successful project, advancing our understanding of the Milky Way Galaxy through 
panoramic imaging of the main constituents of the Interstellar Medium. Observations 
for the CGPS are now complete and the Synthesis Telescope at DRAO \citep{DRAO} 
has returned to proposal-driven mode.

The Dominion Radio Astrophysical Observatory invites astronomers to apply for 
observing time with the DRAO Synthesis Telescope. The DRAO ST provides radio 
observations of atomic hydrogen and radio continuum emission, including the polarized 
signal, with high spatial dynamic range and arcminute resolution. Imaging techniques 
developed for the CGPS have made the telescope into a front-line instrument for 
wide-field imaging, particularly of polarized emission. We will discuss telescope 
characteristics, show examples of data to demonstrate the unique capabilities of the 
ST, and explain where and how to apply for observing time.\end{abstract}

\section{The DRAO Synthesis Telescope}

\begin{figure}[!htb]
\centerline{
\includegraphics[bb = 10 29 610 391,width=0.643\textwidth,clip=]
{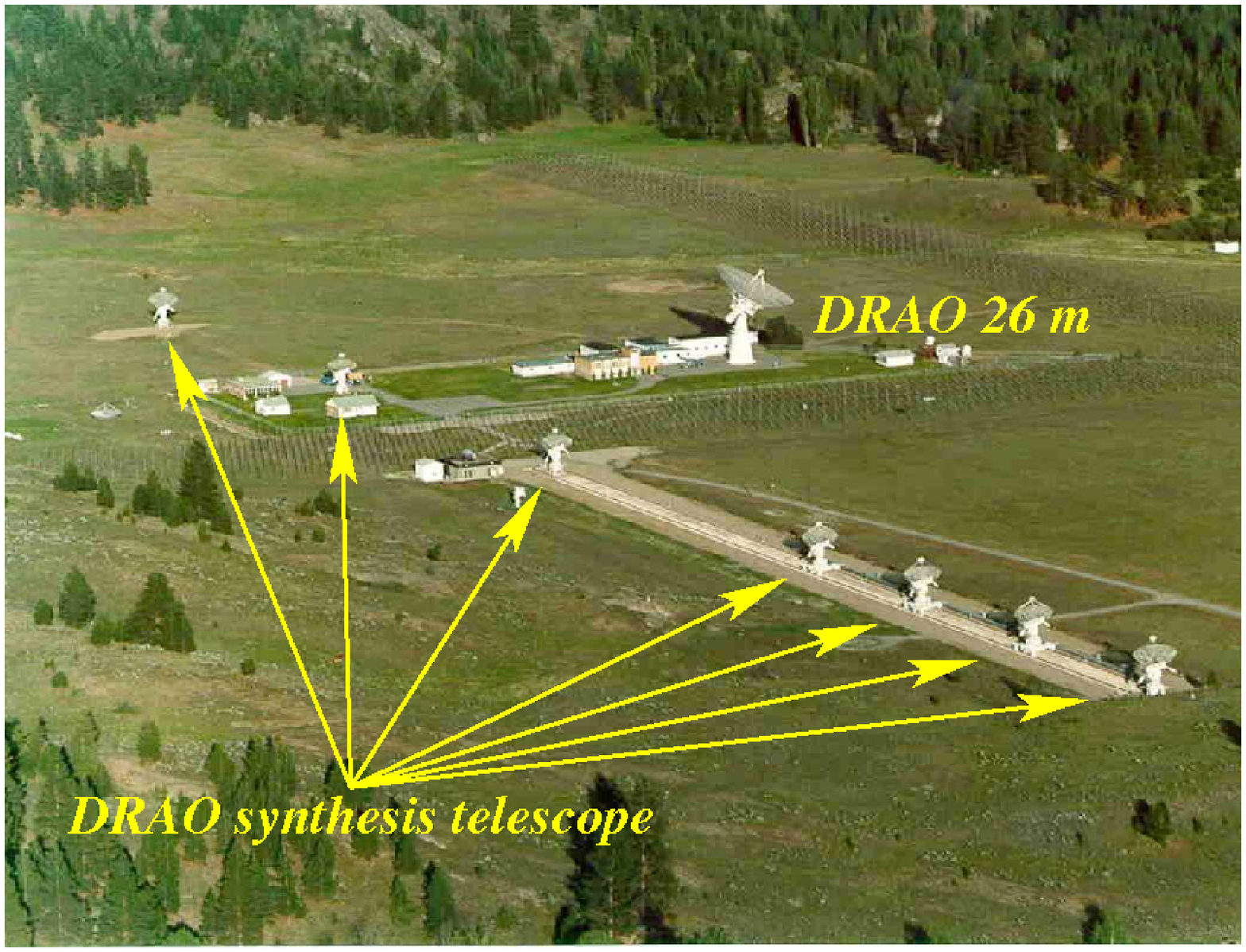}
\includegraphics[bb = 80 60 608 818,width=0.271\textwidth,clip=]
{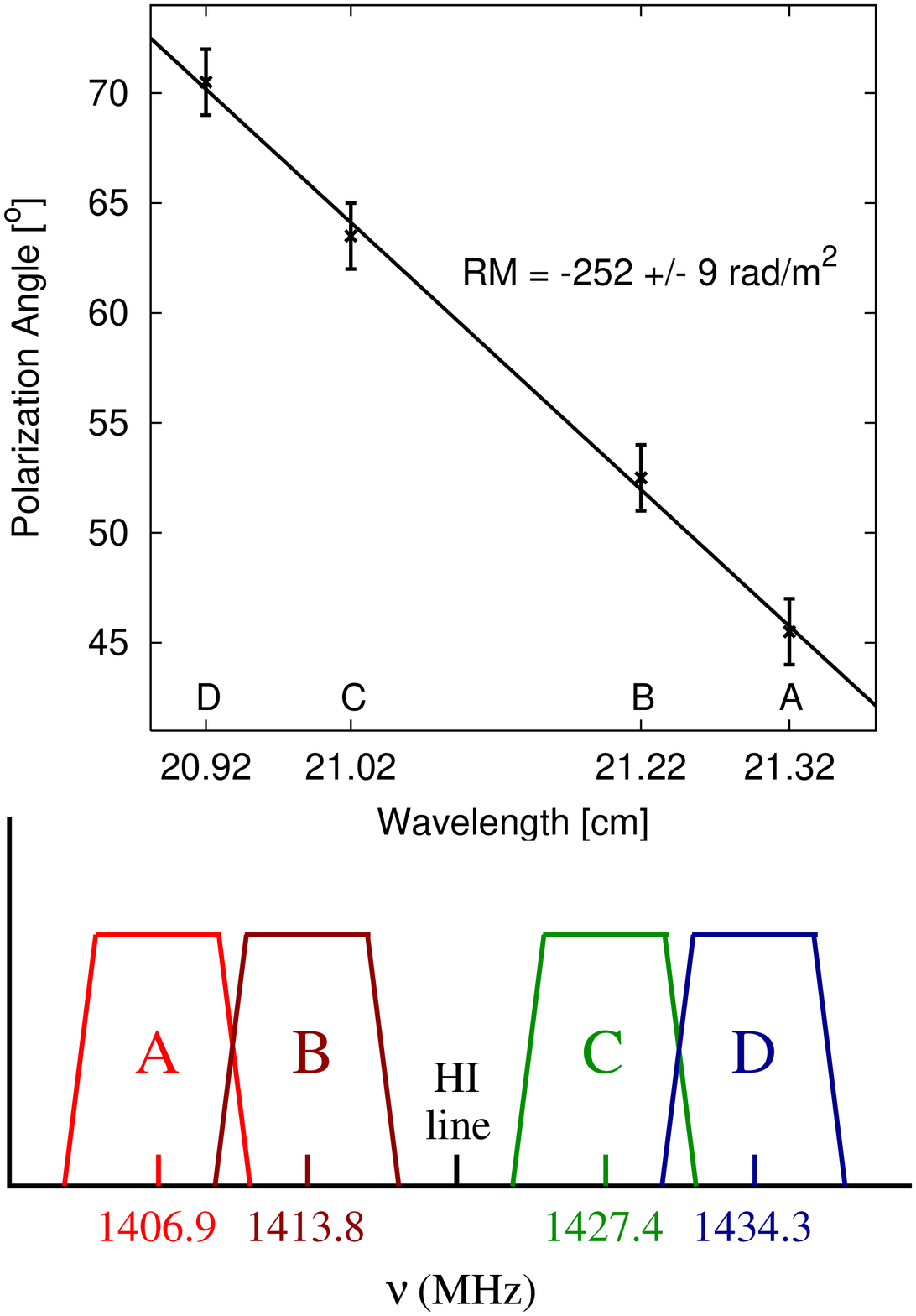}
}
\caption{Aerial picture of the Synthesis Telescope and the 26\,m telescope at
DRAO (left) and the distribution of the four frequency bands around 1420~MHz 
(right).}
\label{fig:st}
\end{figure}

The DRAO Synthesis Telescope \citep[ST, Fig.~\ref{fig:st},][]{DRAO} provides radio 
observations of atomic hydrogen and radio continuum emission at 1420 and 408~MHz, 
including the linearly polarized signal at 1420~MHz, with high spatial dynamic range. 
The ST is a seven element east-west interferometer providing full uv-coverage between 
a maximum spacing of 617\,m and a minimum spacing of 12.9\,m. This translates to a 
spatial frequency coverage from $1\arcmin$ to $45\arcmin$ at 1420~MHz and $2.8\arcmin$ 
to $2.6\deg$ at 408~MHz. The four separate bands (see Fig.~\ref{fig:st} right) available 
around 1420~MHz allow precise determination of rotation measure. Low spatial frequency 
information can be provided by the DRAO 26\,m radio telescope (see Fig.~\ref{fig:st} 
left) upon request. Imaging techniques developed for the CGPS \citep{will99} have made 
the telescope into a front-line instrument for wide-field imaging, particularly of 
polarized emission. Main characteristics of the DRAO Synthesis Telescope are listed in 
Table~\ref{tab:st}.

\begin{table}
\label{tab:st}
\caption{Characteristics of the DRAO Synthesis Telescope}
\centerline{\begin{tabular}{|l|ll|} \hline
 & & \\
Frequencies & 1420~MHz & 408~MHz\\
 & & \\ \hline
 & & \\
Field of View & $2.65\deg$ & $8.22\deg$\\
Angular Resolution & $58\arcsec \times 58\arcsec cosec(\delta)$ & $2.8' \times 2.8' cosec(\delta)$\\
Spatial Frequencies & $1\arcmin$ to $45\arcmin$ & $2.8\arcmin$ to $2.6\deg$ \\
System Temperature & 45~K & 105~K + T$_{\rm sky}$ \\
Continuum Sensitivity & 180~$\mu$Jy/beam & 3.0~mJy/beam \\
Rotation Measure & 4 bands, $\delta(f) = 7.5$\,MHz & \\
Determination & see Fig.~\ref{fig:st} right & \\
Mosaicking Sensitivity & $45\mu$Jy achieved & confusion limited \\
\ion{H}{i} sensitivity & $2.5 B^{-0.5} sin{\delta}$\,K & \\
\ion{H}{i} bandwidths & $B = 0.125, 0.25, 0.5,$ & \\
 & $ 1.0, 2.0, 4.0$\,MHz & \\
 & & \\ \hline
\end{tabular}}
\end{table}

\section{Large Structure Matters!}

\begin{figure}[!htb]
\centerline{
\includegraphics[width=0.364\textwidth]{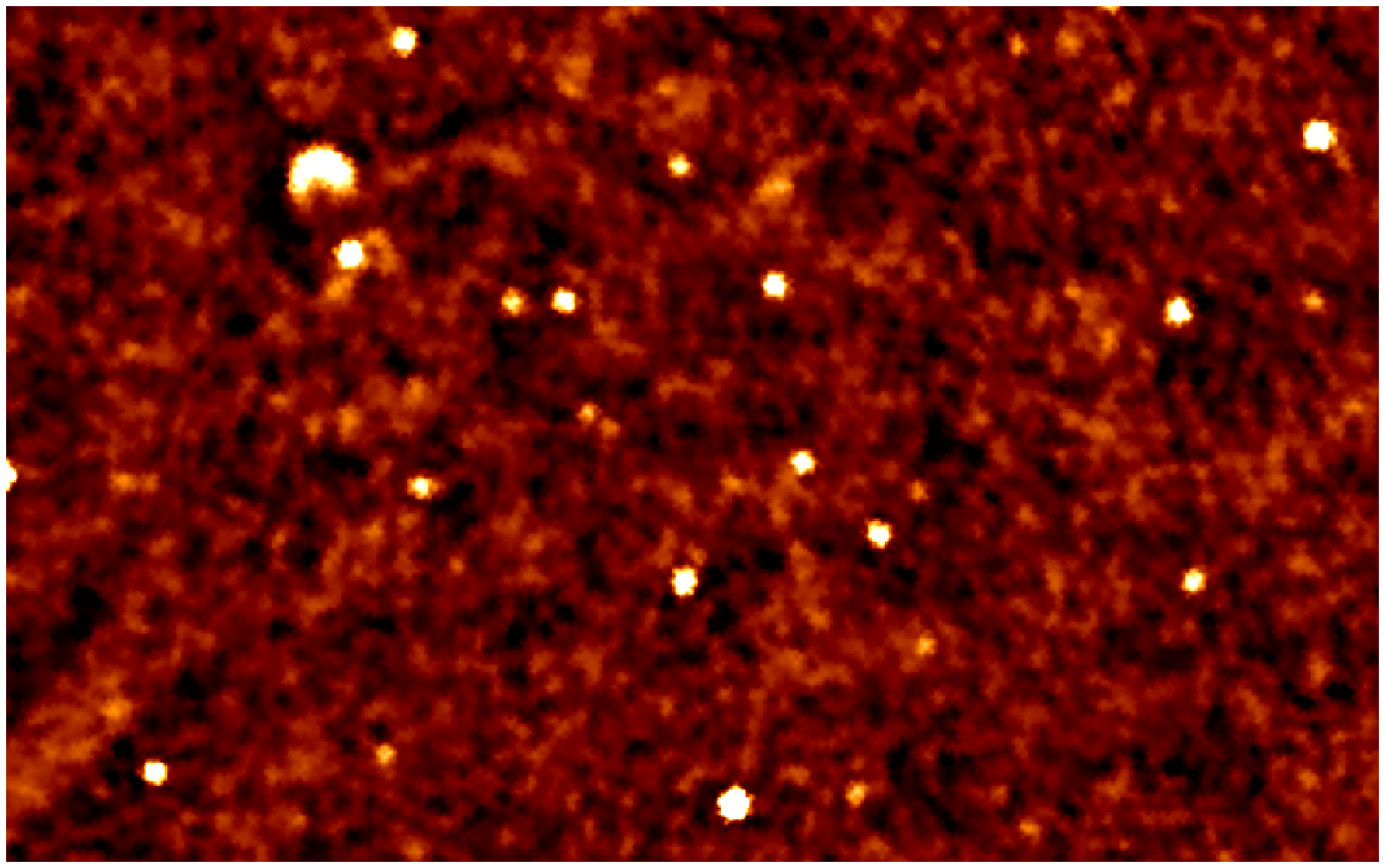}
\includegraphics[width=0.364\textwidth]{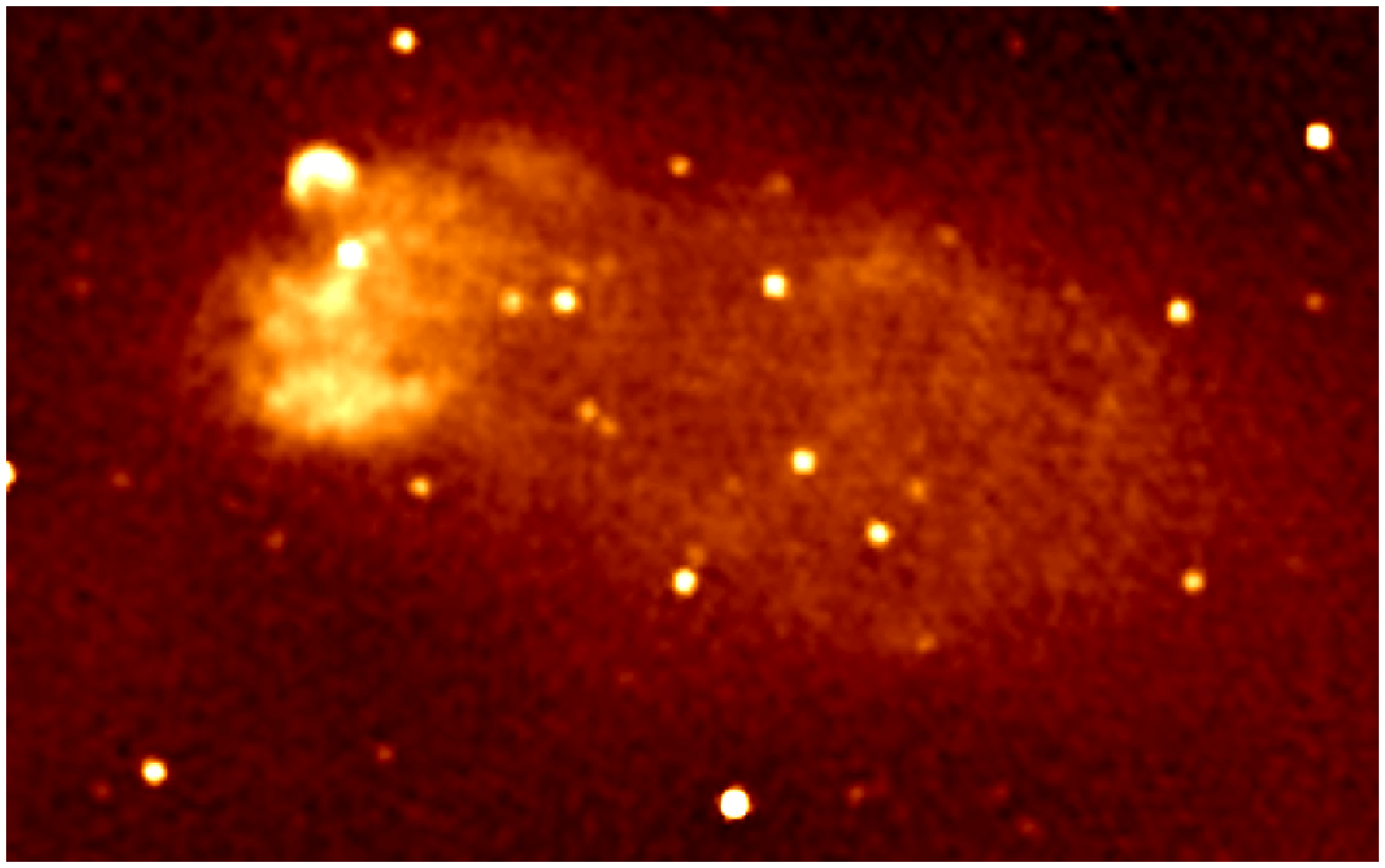}
\includegraphics[bb= 55 55 545 545,width=0.227\textwidth,clip=]{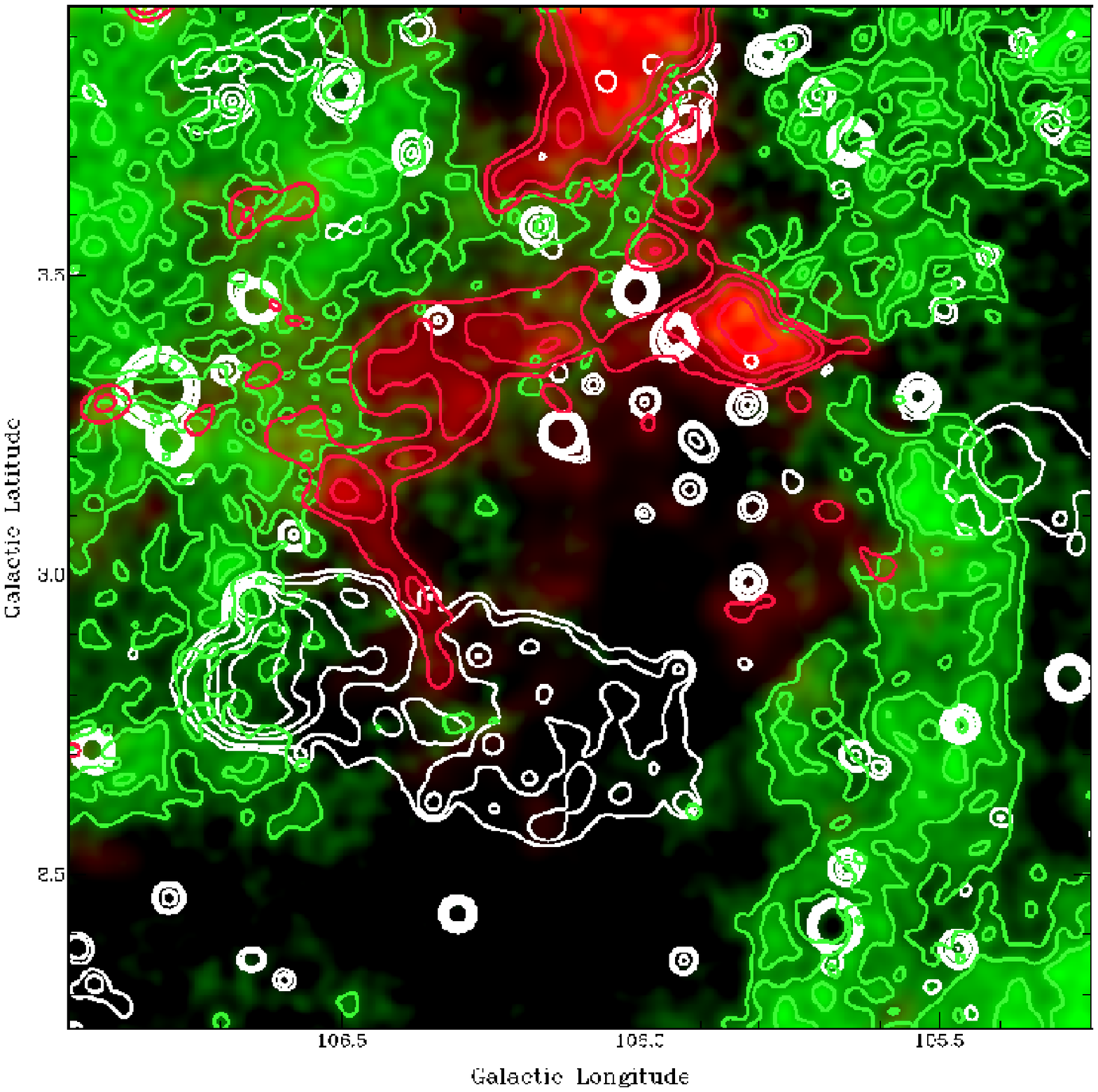}
}
\caption{Left: Stokes {\it I} image from the NVSS. Centre:
Stokes I image of the same area as left taken from the CGPS. Right: Combined
CO (red) and \ion{H}{i} (green) image of the area around the SNR G106.3+2.7,
indicated by the white contours.}
\label{fig:g106}
\end{figure}

Information about the large structure is important to get a complete picture.
The boomerang-shaped object in the top left of Fig.~\ref{fig:g106}
is a pulsar wind nebula and a $\gamma$-ray emitter. The CGPS image 
reveals a large extended object, the associated supernova remnant
G106.3+2.7, which is not visible in the Northern VLA Sky Survey
\citep[NVSS,][]{nvss} image because the VLA cannot detect large structure.

The ``Boomerang'' is a young pulsar wind nebula with a strong mG magnetic field
\citep{kru06}. The environment of this system is
revealed by the right image in Fig.~\ref{fig:g106} of \ion{H}{i} (green) and 
CO (red). The progenitor star, located in the shell of a stellar wind 
bubble, is likely the result of triggered star formation \citep{koth01}.

\section{Polarization Imaging}

\begin{figure}[!htb]
\centerline{
\includegraphics[bb = 20 20 535 410,width=0.7\textwidth]{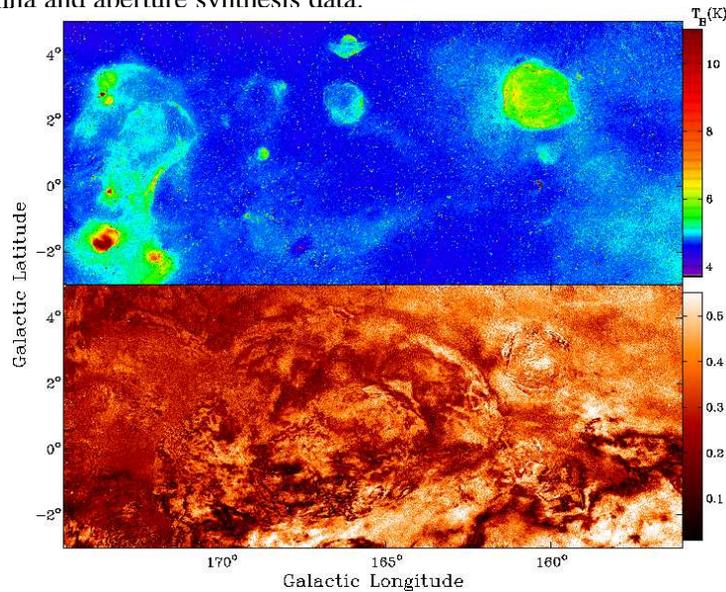}
}
\caption{CGPS Stokes {\it I} image (top) and polarized intensity (bottom) at 1420~MHz of
an area close to the anti-centre region of our Galaxy.}
\label{fig:bubble}
\end{figure}

The CGPS 1420~MHz Stokes {\it I} and polarized intensity images displayed in Fig.~\ref{fig:bubble}
combine data from the DRAO ST with single antenna observations to truly represent all 
structure from the largest scales down to the resolution limit of about $1\arcmin$. 
1420~MHz is close to the ideal frequency for radio polarization imaging, showing both
emission and Faraday rotation features. The DRAO ST is currently {\bf THE} world leading 
radio telescope for imaging linearly polarized emission at that frequency through 
unique capabilities to correct for instrumental polarization \citep{reid08} and 
accurately combine single antenna and aperture synthesis data.

The image in Fig.~\ref{fig:bubble} shows a large stellar wind bubble that was
discovered through its Faraday rotation imprint on the polarized background 
emission. It also may have commenced a star burst in the Outer Galaxy as indicated
by the presence of numerous very young and massive stars and clusters close to
its boundary (Kothes et al., in prep.).

\section{\ion{\bf H}{i} in Nearby Galaxies}

\begin{figure}[!htb]
\centerline{
\includegraphics[bb= 45 20 600 620,width=0.40\textwidth,clip=]{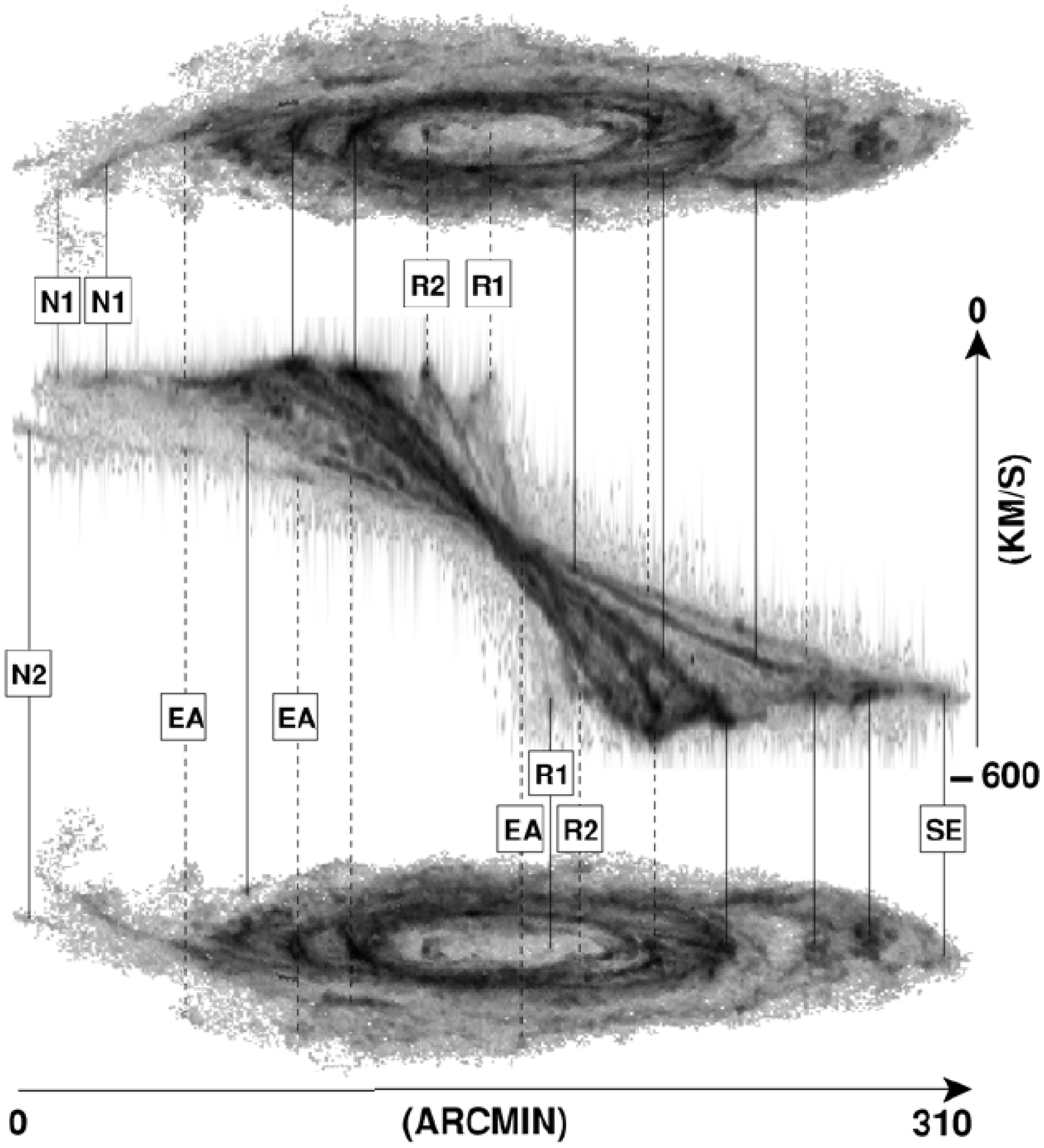}
\includegraphics[bb= 15 13 260 215,width=0.52\textwidth,clip=]{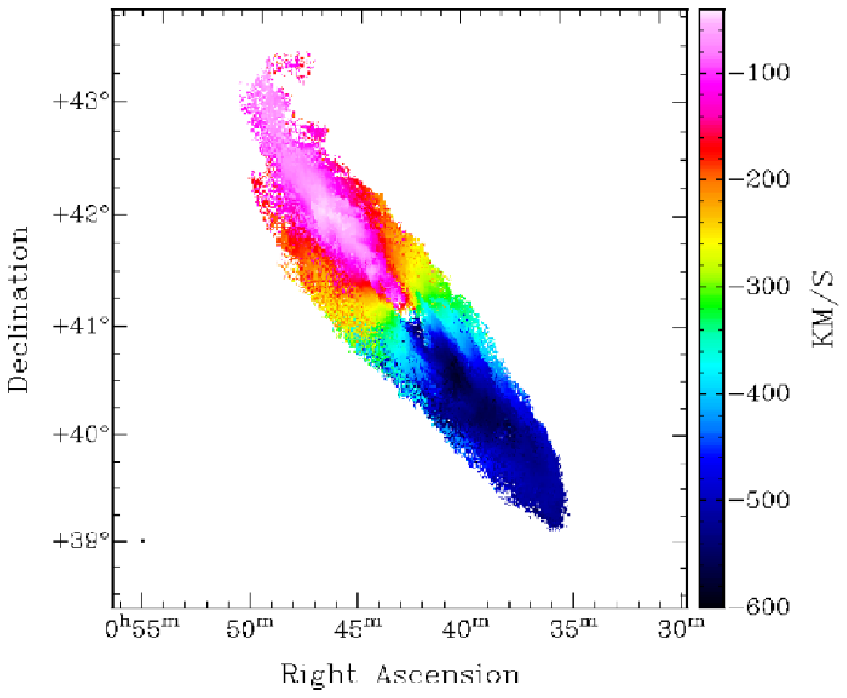}
}
\caption{Integrated \ion{H}{i} images of M\,31 and the projection of the \ion{H}{i} 
data cube onto the 
photometric major axis (left) and the \ion{H}{i} velocity field of M\,31 (right).}
\label{fig:m31}
\end{figure}

The results of new deep \ion{H}{i} observations of the nearby galaxy M\,31 are shown
in Fig.~\ref{fig:m31}. This deep \ion{H}{i} imaging led to the discovery of a faint 
external spiral arm (labeled EA) and new disk 
extremities (S\,E, N\,1, and N\,2) \citep{chem09} as can easily be seen in the 
projection of the data cube onto the photometric major axis (Fig.~\ref{fig:m31}, 
left).

\section{Sensitivity Through Deep Integration}

\begin{figure}[!htb]
\centerline{
\includegraphics[width=0.89\textwidth]{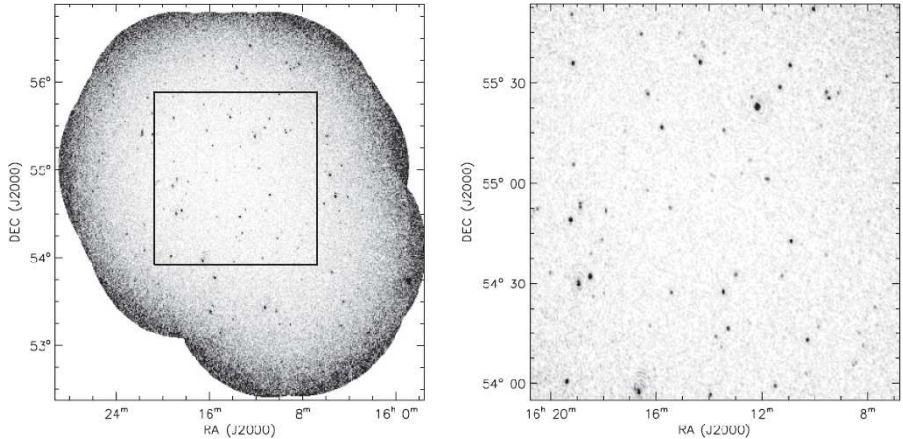}
}
\caption{A 40-pointing mosaic of polarized intensity at 1420~MHz of the ELAIS N1 field. 
The images show polarized intensity with a grayscale from 0 to 1 mJy/beam for the whole 
area (left) and for the central region (right).}
\label{fig:deep}
\end{figure}

A mosaic of 40 pointings was observed towards the ELAIS~N1 field to achieve the most
sensitive wide-field image of linear polarization at 1420~MHz ever observed (Fig.~\ref{fig:deep}).
A sensitivity for Stokes {\it I} of 55\,$\mu$Jy/beam and for Stokes {\it Q} and {\it U} 
of 
45\,$\mu$Jy/beam was achieved. Further observations of the ELAIS~N1 area are underway 
to reach even
higher sensitivity of around 30\,$\mu$Jy/beam for Stokes {\it Q} and {\it U}.

The images shown in Fig.~\ref{fig:deep} are the basis for new results on counts 
of faint polarized extragalactic sources. A higher degree of fractional polarization
was found for sources with fainter Stokes {\it I} than for brighter sources. This highly 
polarized faint source population seems to be dominated by radio galaxies 
\citep{Grant_2010}.

\section{Imaging Galactic \ion{\bf H}{i} at Arcminute Resolution}

\begin{figure}[!htb]
\centerline{
\includegraphics[width=0.89\textwidth]{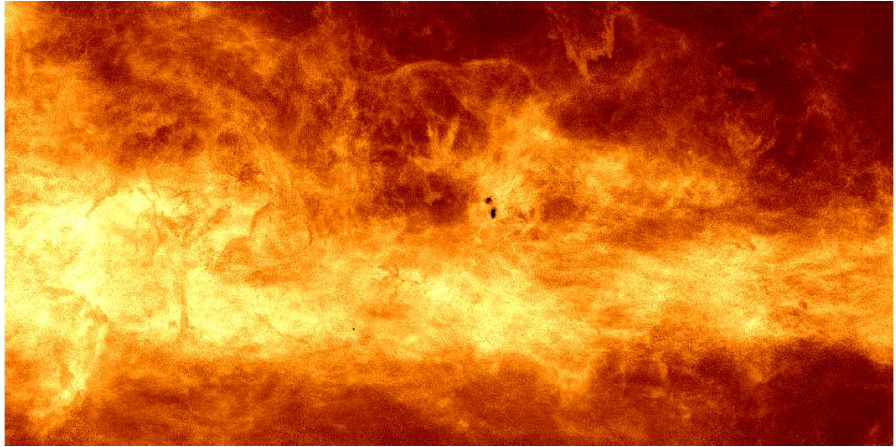}
}
\caption{Image of hydrogen emission over an area $20^\circ \times 9^\circ$ in size 
centered at $\ell = 135^\circ, b = 1^\circ$, showing \ion{H}{i} gas at Perseus arm velocities.}
\label{figu:hi}
\end{figure}

Fig.~\ref{figu:hi} shows an image of \ion{H}{i} emission towards the massive W3/4/5 
\ion{H}{ii} region
complex taken from the Canadian Galactic Plane Survey Data base. Data from the DRAO 
Synthesis Telescope and 26-m Telescope have been combined. The data from the Canadian 
Galactic Plane Survey are available at the Canadian Astronomical Data Centre at:
\centerline{
{http://cadc.hia.nrc.ca/cgps}
}

In Fig.~\ref{figu:hi} filamentary
structure is seen down to the resolution limit of the CGPS. The 
narrow, dark shadows reveal cold foreground HI absorbing emission from brighter and
warmer, more distant gas. These \ion{H}{i} self absorption structures (HISA) are very common
in the CGPS. This is cold atomic hydrogen recently compressed by the passage of a 
spiral shock on its way to forming molecules. We are seeing the first step in forming 
stars \citep{Gibson2005a}.

\section{Observing with the DRAO Synthesis Telescope}

There is currently no proposal deadline for the DRAO Synthesis Telescope, proposals are 
received and refereed throughout the year. If you wish to learn more about observing at 
DRAO, please contact our Operations Manager at:
\begin{center}
Andrew.Gray@nrc-cnrc.gc.ca
\end{center}

\acknowledgements

The Dominion Radio Astrophysical Observatory is a National Facility
operated by the National Research Council. The Canadian Galactic
Plane Survey is a Canadian project with international partners, and is
supported by the Natural Sciences and Engineering Research Council
(NSERC).

\bibliography{master}

\end{document}